\documentstyle[epsf,epsfig,eqsecnum,aps,amssymb,multicol]{revtex}

\begin{document}
\draft \preprint{}
\title{Scanning SQUID Microscope Study of Vortex Polygons and Shells in Weak Pinning Disks of an Amorphous Superconducting Film}
\author{Nobuhito Kokubo$^1$, Satoru Okayasu$^2$, Akinobu Kanda$^3$, and Bunju Shinozaki$^4$\\
}

\address{$^1$ Center for Research and Advancement in Higher Education, Kyushu University, Fukuoka 819-0395, Japan\\}
\address{$^2$ Advanced Science Research Center, Japan Atomic Research Institute, Ibaraki 319-1195, Japan\\}
\address{$^3$ Institute of Physics, University of Tsukuba, Tsukuba 305-8571, Japan\\}
\address{$^4$ Department of Physics, Kyushu University, Fukuoka 812-8581, Japan\\}

\date{\today}
\maketitle
\begin{abstract}
Direct observation of vortices by the scanning SQUID microscopy was
made on large mesoscopic disks of an amorphous MoGe thin film. Owing
to the weak pinning nature of the amorphous film, vortices are able
to form geometry induced, (quasi-)symmetric configurations of
polygons and concentric shells in the large disks. Systematic
measurements made on selected disks allow us to trace not only how
the vortex pattern evolves with magnetic field, but also how the
vortex polygons change in size and rotate with respect to the disk
center. The results are in good agreement with theoretical
considerations for mesoscopic disks with sufficiently large
diameter. A series of vortex images obtained in a disk with a
pinning site reveals a unique line symmetry in vortex
configurations, resulting in modifications of the shell filling rule
and the magic number.
\end{abstract}



\renewcommand{\theequation}{\arabic{equation}}
\begin{multicols}{2}
\section*{I. Introduction}
Tiny superconductors, accommodating only a few quantized magnetic
flux lines, display a great variety of vortex structure different
from the Abrikozov vortex lattice in bulk superconductors
\cite{Abrikosov1957}. Examples include vortex polygons and
concentric vortex rings called "vortex shells", which are stable
configurations of the repulsive flux lines under the geometrical
confinement via the influence of the screening current flowing along
the sample edge. The issue of vortex matter confined into small
superconductors has been studied theoretically and numerically for
many years, focusing mainly on how the vortices are distributed in
disk
\cite{BuzdinPL1994}\cite{Venegas1998}\cite{PalaciosPRL2000}\cite{BaelusPRB2004}\cite{CabralPRB2004}\cite{MiskoPRB2007},
square \cite{ChibotaruNature2000}, and triangle shaped small
superconductors \cite{ChibotaruPRL2001}. Several studies reveal the
field evolutions of the vortex structure in small superconductors,
and argue which vortex configurations are energetically favorable
and how the transition between different vortex states occurs. The
obtained rule of shell filling and magic number configurations for
consecutive new shells are relevant to phenomena observed in other
systems, including the puzzling nucleation of vortices observed in
rotating condensates of superfluid $^4$He \cite{YarmchuckJLT1982}
and cold dilute alkali-metal gases \cite{MadisonPRL2000}.

The experimental investigations for vortex states in small
superconductors were initiated by Hall magnetometer measurements
\cite{GeimNature1997}\cite{GeimPRL2000}, followed by a
multiple-small tunneling junction measurement \cite{KandaPRL2003}.
Kinks in the magnetization or jumps in the tunneling spectra mark
transitions of vortex states in small superconducting dots. They
evidence changes of vorticity $L$ (the number of vortices), not
actual distributions of vortices in the dots.

The visualization of vortices in small superconductors was reported
earlier in studies of the scanning superconducting quantum
interference device (SQUID) microscopy
\cite{HataPhysC2003}\cite{NishioPhysC2004} and later in studies of
the scanning Hall probe microscopy
\cite{NishioPRB2008}\cite{ConnollyEPL2009}, showing reasonable
images of vortex configurations for $L$ up to 6. More vortices have
been imaged in recent Bitter decoration studies as patterns of
magnetic particles deposited over micrometer-sized dots of Nb films
\cite{GrigorievaPRL2006}\cite{ZaoPRB2008} and Nb
mesas\cite{GrigorievaPRL2007}. The experiments reveal the rule of
the shell filling with magic numbers for $L$ up to 40 by combining
vortex configurations visualized over many disks at different
magnetic fields, although some features influenced by the bulk
pinning and/or the roughness of the disks are involved.


The scanning SQUID microscope technique, which we employed in this
study, is an alternative and complement method to visualize vortices
in the superconducting dots. By scanning a small pick-up loop over
the dots, one can magnetically image the vortices without damaging
the samples. This allows us to study systematically how the vortex
configuration evolves with magnetic field in a specific sample.
While the scanning SQUID microscope has attained the best magnetic
sensitivity (better than 5 $\mu \Phi_0/\sqrt{\texttt{Hz}}$), the
spatial resolution is limited typically to a few micrometer due to
the size of the pick-up loop ($\sim$ 10 $\mu$m)
\cite{KirtleyARL1995}. This implies that the observable vortex
density is low ($<$ 100 $\mu$T) and the sample becomes much larger
than the loop size. The earlier studies were made on YBCO
\cite{HataPhysC2003} and Nb dots \cite{NishioPhysC2004} with $\sim$
50 $\mu$m in size, which is decades larger than the penetration
depth $\lambda$ or the coherence length $\xi$ of the superconducting
materials. In such large dots, the effect of the geometrical
confinement is weak and (sparse) vortices form likely disordered
configurations due to the dominant influence of the bulk pinning in
the superconducting materials
\cite{HataPhysC2003}\cite{NishioPhysC2004}.


If the influence of the bulk pinning is reduced by using weak
pinning, superconducting materials, the situation can be different:
the geometrical confinement dominates over the bulk pinning.
Moreover, when the dots become thin in thickness $d$ $( <
2\lambda)$, the interaction between vortices becomes long ranged and
it decays slowly with the vortex-vortex spacing $r$ as $\propto 1/r$
for $r > 2\lambda^2/d$ \cite{PearlAPL1964}, in contrast to the short
ranged interaction characterized by the exponential decay $\propto
e^{-r/\lambda}$ for $r
> \lambda$ in the bulk superconductors. Thus, even in the large dots
the interplay between the vortex-vortex interaction and the
geometrical confinement may lead to geometry-induced, symmetric
configurations of the sparse vortices in the presence of the weak
bulk pinning. Here, we report on the direct observation of vortices
in large thin disks of an amorphous ($\alpha$) MoGe film by the
scanning SQUID microscopy. Due to the weak bulk pinning of the
amorphous film, we are able to observe (quasi-) symmetric
configurations of polygons and concentric shells of vortices for $L$
up to 19. The results illustrate not only how the vortex
configuration evolves with the magnetic field, but also how the
vortex polygons change in size and rotate with respect to the disk
center. We present vortex images obtained in a disk with a pinning
site and discuss how vortex configurations are altered by a pinned
vortex.

This paper is outlined as follows: After describing experimental
details in Sec. II, we present vortex polygon states observed for
small vorticities up to $L$= 5 in Sec. III. In Sec. IV we present
field evolutions of vortex shells observed for large vorticities
$L\geq$ 6. The influence of a pinning site on vortex configurations
is discussed in Sec. V. In Sec. VI comparison with other experiment
is made. Our summary is given in Sec. VII.

\section*{II. Experimental}
We used a commercial scanning SQUID microscope (SQM-2000, SII
Nanotechnology) with a dc SQUID magnetometer made of
Nb/Al-AlO$_x$/Nb Josephson junctions and an inductively coupled,
pick-up loop of a Nb film \cite{NishioPhysC2004}.  The pick-up loop
had 10 $\mu$m in diameter with 2 $\mu$m line width.  The dc SQUID
and the pick-up loop were integrated on a square Si chip of 3
$\times$ 3 mm$^2$ in size, which was mounted on a phosphor bronze
cantilever. The sensor chip was tilted with respect to the sample
stage by a shallow angle of $\sim$ 10 degrees and a corner of the
chip was softly in contact with the sample surface, in order to keep
a short distance ($\sim$ 5 $\mu$m at the minimum) between the loop
and the sample surface. This ensures safe operation of the scanning
SQUID microscope with reasonably high spatial resolutions ($\sim$ 4
$\mu$m). The whole assembly including the sensor chip and the sample
stage was surround by a $\mu$-metal shield, resulting in a residual
magnetic field (ambient field) of $\sim$ 1 $\mu$T. A coil wound in
our sample stage allowed us to apply small magnetic field $H$
perpendicular to the sample. We scanned the sample stage in XY
directions by dc stepper motors. In order to minimize the possible
mechanical hysteresis effect, we took one scan direction from the
left side to the right side of the image. The scanning speed was
about 5 pixels/s and this resulted in an image of 64$\times$64
pixels by typically 15 minutes. The temperature for the SQUID sensor
was 2 - 3 K, while that for the sample stage was varied.

We prepared weak pinning disks of an $\alpha$-Mo$_{x}$Ge$_{1-x}$ ($x
\approx$ 78 $\%$) thin film with a lithographic technique. After
preparing a silicon substrate covered with a patterned-resist film,
we sputtered the $\alpha$-MoGe film with 0.22 $\mu$m thickness on
top. Employing the lift-off technique, we patterned the film into
arrays of disks with four different diameters $D_d$ of 20 $\mu$m, 34
$\mu$m, 56 $\mu$m and 106 $\mu$m. The spacing between the disks was
chosen to be large enough ($>$ 200 $\mu$m), so that the interdisk
coupling can be excluded. In order to reduce possible damage during
the scanning, all the disks were covered with 0.1 $\mu$m thick
SiO$_2$ film as a protective layer. The material parameters relevant
to this study are as follows: the superconducting transition
temperature $T_c$ is 6.0 K, the coherence length $\xi(0)$ at $T = 0$
is 4.5 nm, and the penetration depth $\lambda (0)$ at $T = 0$ is
0.58 $\mu$m ($ \gg \xi(0)$) \cite{KesPRB1983}. Since the thickness
is smaller than $\lambda(0)$, the penetration depth is expanded to
the effective one given by $\Lambda_{eff}(0)=2\lambda^2(0)/d$ $(\gg
\lambda(0))$. For the present film $\Lambda_{eff}(0) \sim$ 3.1
$\mu$m. Normalizing the disk size by the effective penetration
depth, we find  $D_d \approx$ 6-35 $\Lambda_{eff}(0)$, indicating
large mesoscopic disks.

In weak pinning materials, the distance $z_0$ between the pick-up
loop and the sample surface is an important parameter to be
controlled in order to obtain clear images of vortices. As
demonstrated by Plourde \emph{et al.} \cite{PlourdePRB2002}, the
vortices may be dragged and/or shifted in positions by the pick-up
loop during the scanning. This coupled motion of vortices can be
reduced when $z_0$ is increased by lowering the sample stage from
the contact point with the sensor chip. This procedure, however,
results in a smeared image of vortices due to the spread of the
magnetic flux above the sample surface \cite{KoganPRB2003}. We could
not resolve individual vortices when $z_0$ became more than the loop
diameter ($>10$ $\mu$m). Hence, we set $z_0 \sim$ 7 $\mu$m as a
typical distance. All the images presented in this study were taken
in the field-cool measurement, where the magnetic field was applied
in some temperatures ($\sim$10 K) above $T_c$, followed by cooling
the samples in the magnetic field to temperatures ($\sim$ 3-5 K)
below $T_c$ where scanning SQUID measurements were done. This
procedure results in the equilibrium distribution of vortices.
Because of the possible motion of vortices during the scanning, each
image was taken in a single scan made on a fresh vortex state
prepared by the field-cool procedure. No image taken by repeated
scans is presented.

\section*{III. Vortex configurations for small vorticities}

First we focus on vortex configurations observed for small
vorticities up to 5. Figs. 1(a)-1(f) show selected vortex images
observed in a 34 $\mu$m disk. A color bar indicates the magnitude of
the magnetic flux in the pick-up loop. These images display clearly
individual vortices with reasonable spatial and magnetic resolutions
\cite{rmsnoise}. The imaged vortex size is larger than
2$\Lambda_{eff} (\sim$ 7 $\mu$m) due to defocusing effect
originating from the large distance $z_0 (\sim$ 7 $\mu$m) between
the loop and the sample surface. Details of the field profile of
vortices have been reported in Refs.
\cite{PearlAPL1964}\cite{TafriPRL2004}\cite{NishioKokuboPRB2008}. We
find that the vortices form symmetric configurations with respect to
the disk center.  The vortex pattern evolves with field as follows;
After the Meissner state where the magnetic field is expelled from
the disk, a single vortex appears in the disk and it sits at the
disk center (Fig. 1(a)). Then, two vortices are located
symmetrically with respect to the disk center (Fig. 1(b)), followed
by the formation of a triangle (Fig. 1(c)), a square (Fig. 1(d)) and
a pentagon (Fig. 1(e)) of vortices in the disk.
Although some distortion is visible for the square, these symmetric
configurations are nearly identical to those reported in the Bitter
decoration \cite{GrigorievaPRL2006} and the theoretical studies
\cite{BaelusPRB2004}\cite{CabralPRB2004}\cite{MiskoPRB2007}.

\begin{figure}
\begin{center}
\epsfig{file=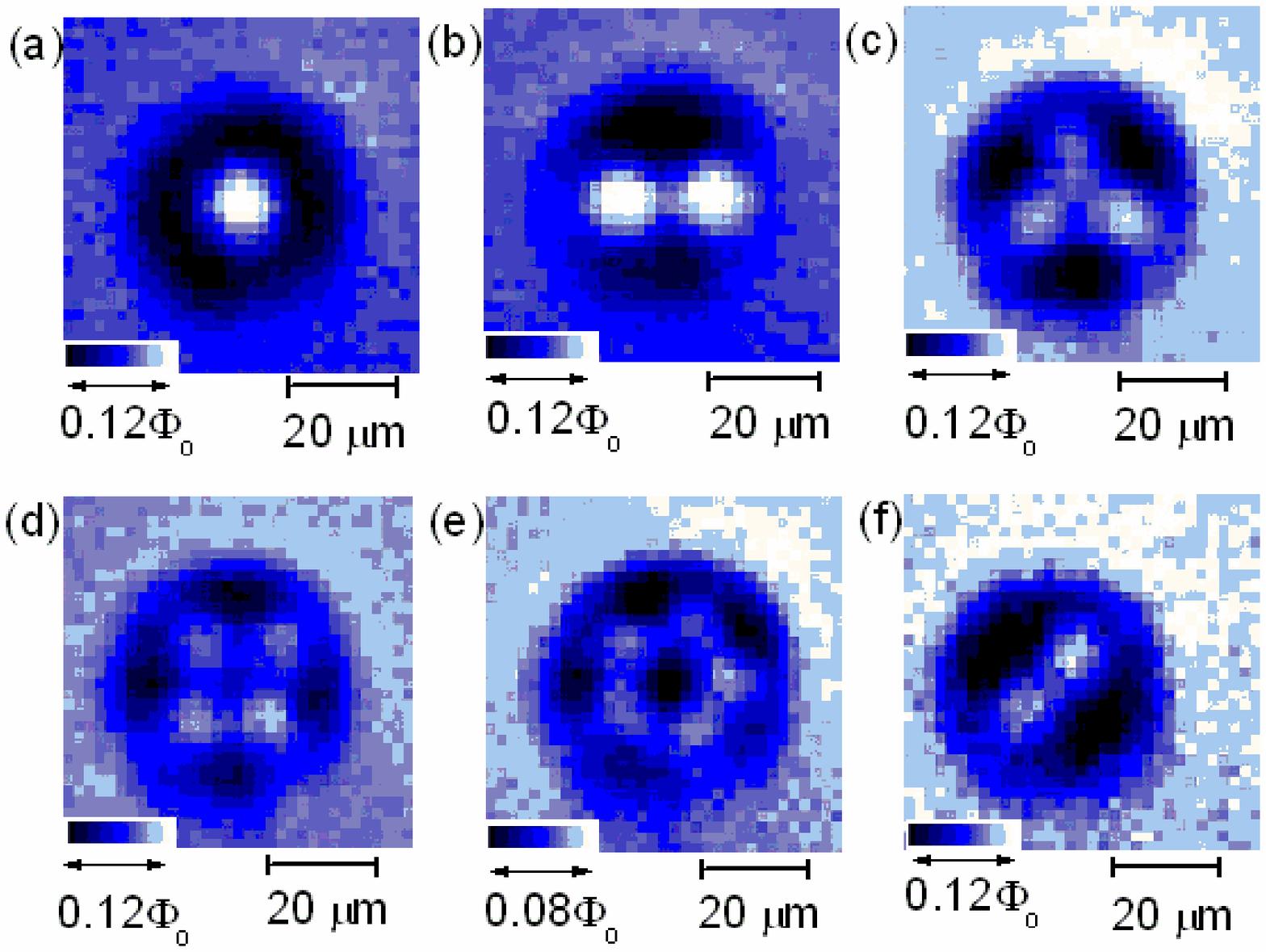, width=7 cm} \vspace{0.1cm} \caption{
Scanning SQUID microscope images of vortices in an amorphous MoGe
disk with 34 $\mu$m diameter for vorticities $L$= 1-5 obtained after
the disk was cooled to 3.2 K in different magnetic fields. (a) $L$=
1 at 8.0 $\mu$T; (b) $L$= 2 at 10 $\mu$T; (c) $L$= 3 at 16 $\mu$T;
(d) $L$= 4 at 18 $\mu$T; (e) $L$= 5 at 21 $\mu$T. (f) An additional
image of the vortex pair at 14 $\mu$T. All images are the same in
size of 66*66 $\mu$m$^2$. Color bars indicate the magnitude of the
magnetic flux detected in the pick-up loop. }
\end{center}
\end{figure}

An important remark for the observation is that some of the vortex
patterns rotate with respect to the disk center. This feature is
clearly visible in the images of the vortex pair shown in Figs. 1
(b) and 1(f). In Fig. 1(b), the vortex pair is aligned nearly with
the horizontal direction of the image, while in Fig. 1 (f) it
rotates by $\sim$ 45$^\circ$ with respect to the disk center. Other
orientation of the pair is observed in other scan, indicating the
rotational degree of freedom in the vortex configuration
\cite{rotation}. Similar rotation is observed for the triangle
pattern, but not clearly for the square and the pentagon patterns.

\begin{figure}
\begin{center} \epsfig{file=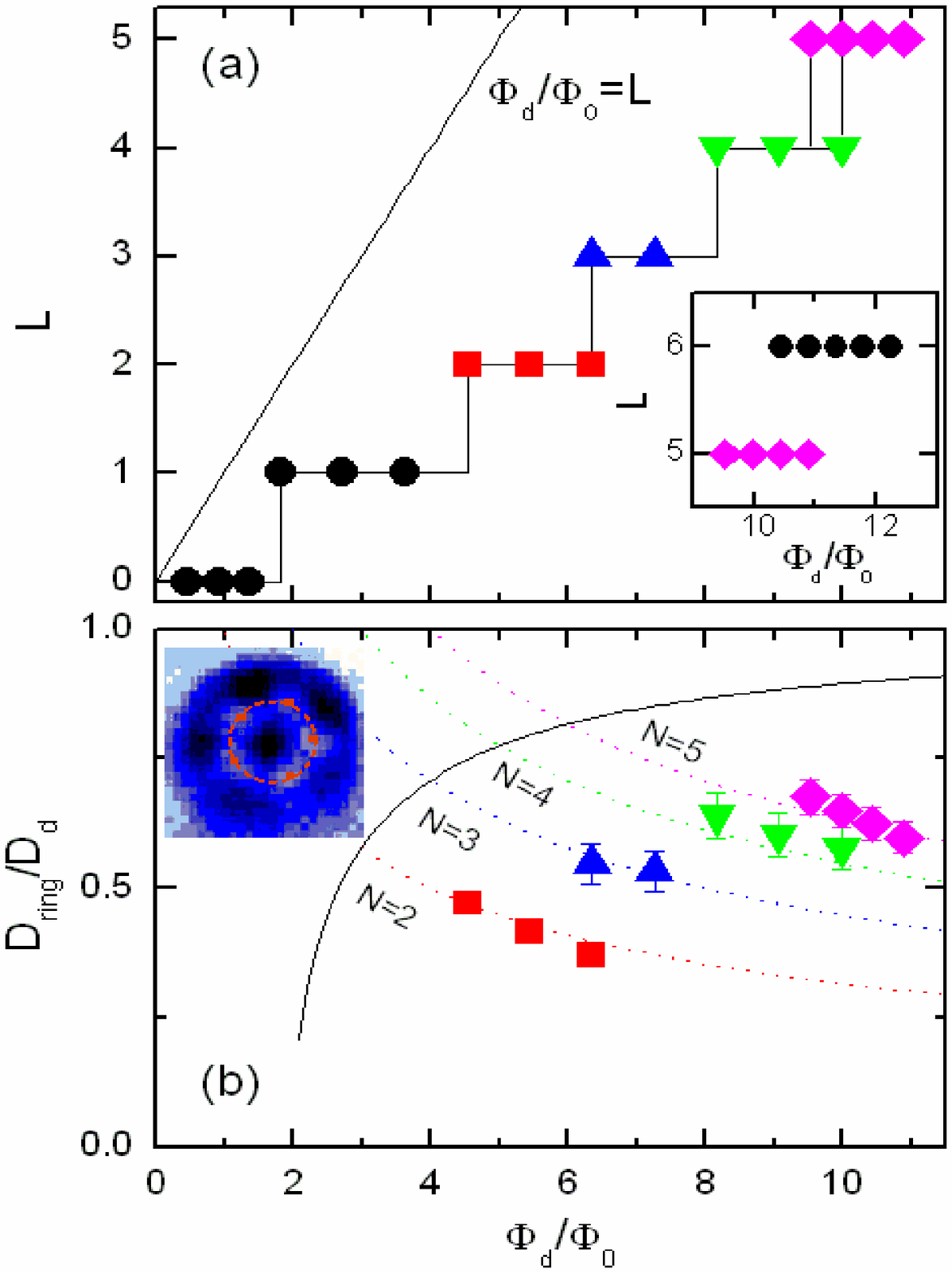,width=8cm} \vspace{0.1cm} \caption{ (a) Vorticity vs. normalized magnetic flux in
the 34 $\mu$m disk in 3.2 K. A solid line represents a condition of
$\Phi_d/\Phi_0=L$. The inset to (a) shows the results around the
transition between $L$= 5 and 6 obtained by repeated field-cool
measurements. (b) The normalized diameter of the polygon ring vs.
the normalized magnetic flux. The dotted curves represent the
theoretical results for polygon sizes with different $N$= 2-5 as
denoted. The solid curve indicates the upper limit of the polygon
size. The inset to (b) illustrates how we determine the diameter
$D_{ring}$ of the polygon ring.}
\end{center}
\end{figure}

Next we show how the number of vortices evolves with applied
magnetic field. Counting the number of vortices in the images taken
at various magnetic fields, we plot the results in Fig. 2(a). Here,
instead of the field strength $H$, we use the magnetic flux
$\Phi_{d}(=\mu_0H\pi(D_d/2)^2)$ in the disk normalized by the flux
quantum $\Phi_0$. As observed, the vorticity $L$ shows a stair case
behavior with field:
After the Meissner state ($L=$ 0), the single vortex state ($L=$ 1)
appears at $\Phi_{d}/\Phi_0\approx$ 2, and it continues until
$\Phi_{d}/\Phi_0\approx$ 4.5, above which the vortex pair state
($L=$ 2) appears. The vortex triangle ($L=$ 3), square ($L=$ 4) and
pentagon ($L=$ 5) states appear at $\Phi_{d}/\Phi_0\approx$ 6.4, 8.2
and 9.5, respectively. We note that the stair case behavior is well
below the condition of $\Phi_d/\Phi_0=L$ where the magnetic field
penetrates fully the disk and this is represented by a solid line.
The difference from the line becomes large on increasing field. At
$\Phi_{d}/\Phi_0 =$ 8, for instance, $\Phi_{d}/\Phi_0-L $= 4, which
is half of $\Phi_{d}/\Phi_0$. This indicates a large diamagnetic
response of the disk and it implies the presence of the large
screening current near the disk edge.

The influence of the screening current upon vortex configurations is
found by considering how the size of vortex polygons changes with
magnetic field. As exemplified in the inset to Fig. 2(b), we draw a
circle circumscribing the positions of vortices and estimate the
diameter $D_{ring}$ of the polygon ring. The results for this
analysis made at different fields and different vorticities (down to
$L$= 2) are summarized in the main panel of Fig. 2(b). As observed,
$D_{ring}$ varies largely with $L$. For example, $D_{ring}$ at $L$=
2 is roughly 40 $\%$ of the disk diameter, while that at $L$= 5
becomes expanded to roughly 65 $\%$ of the disk diameter. Focusing
on the results at each vorticity, we find $D_{ring}$ decreases with
increasing field. The shrinkage of $D_{ring}$ is observed to be
5-10$\%$ of the disk diameter, depending on $L$.

The obtained results are in excellent agreement with the theoretical
considerations for vortex polygons confined by the screening current
near the disk edge \cite{BuzdinPL1994}\cite{CabralPRB2004}. In the
London limit, the diameter of a vortex polygon at a given magnetic
field is written as
\begin{equation}
D_{ring}=D_d\sqrt{\frac{N-1}{\Phi_d/\Phi_0}},
\end{equation}
where $N$ is the number of vortices on the polygon ring. The
behaviors for $N$= 2-5 are represented by dotted curves in Fig.
2(b). Our results for $L=N=$ 2-5 follow nicely those curves without
any adjustable parameters. Moreover, they are satisfactory for the
condition that the polygon size should be below the maximum diameter
above which the formation of the polygon ring is not stable. This is
represented by a solid curve and it follows the condition
\cite{CabralPRB2004} given by
\begin{equation}
D_{ring}=D_d\sqrt{1-\frac{2}{\Phi_d/\Phi_0}}.
\end{equation}
Thus, we are convinced that the vortex configurations observed in
the 34 $\mu$m disk are dominantly influenced by the geometrical
confinement via the screening current near the disk edge. We note,
to the best of our knowledge, that these quantitative agreements,
together with the rotation of the vortex patterns in the same disk,
are not reported in other experimental studies so far.

\section*{IV. Vortex configurations for large vorticities}

For vorticities more than 6, we observe the formation of concentric
shell rings of vortices. Since the vortices become dense and their
magnetic images are overlapped each other, for clarity, we present
here the results obtained in larger disks. Figs. 3(a)-3(c) show
vortex images for $L$= 6-8 observed in a disk with 56 $\mu$m
diameter. One can clearly see the shell formation of vortices. In
the image of Fig. 3(a), for example, a vortex sits at the center of
the disk, while the others form a ring around. Using the standard
notation, this configuration is referred to as (1,5), meaning that
the inner shell has one vortex and the outer shell has five
vortices. After the (1,5) configuration, we observe (1,6) and (1,7)
configurations, which are shown in Figs. 3(b) and 3(c),
respectively. Namely, the increase of vorticity in the disk results
in the growth of the outer shell ring.
\begin{figure}
\begin{center}
\epsfig{file=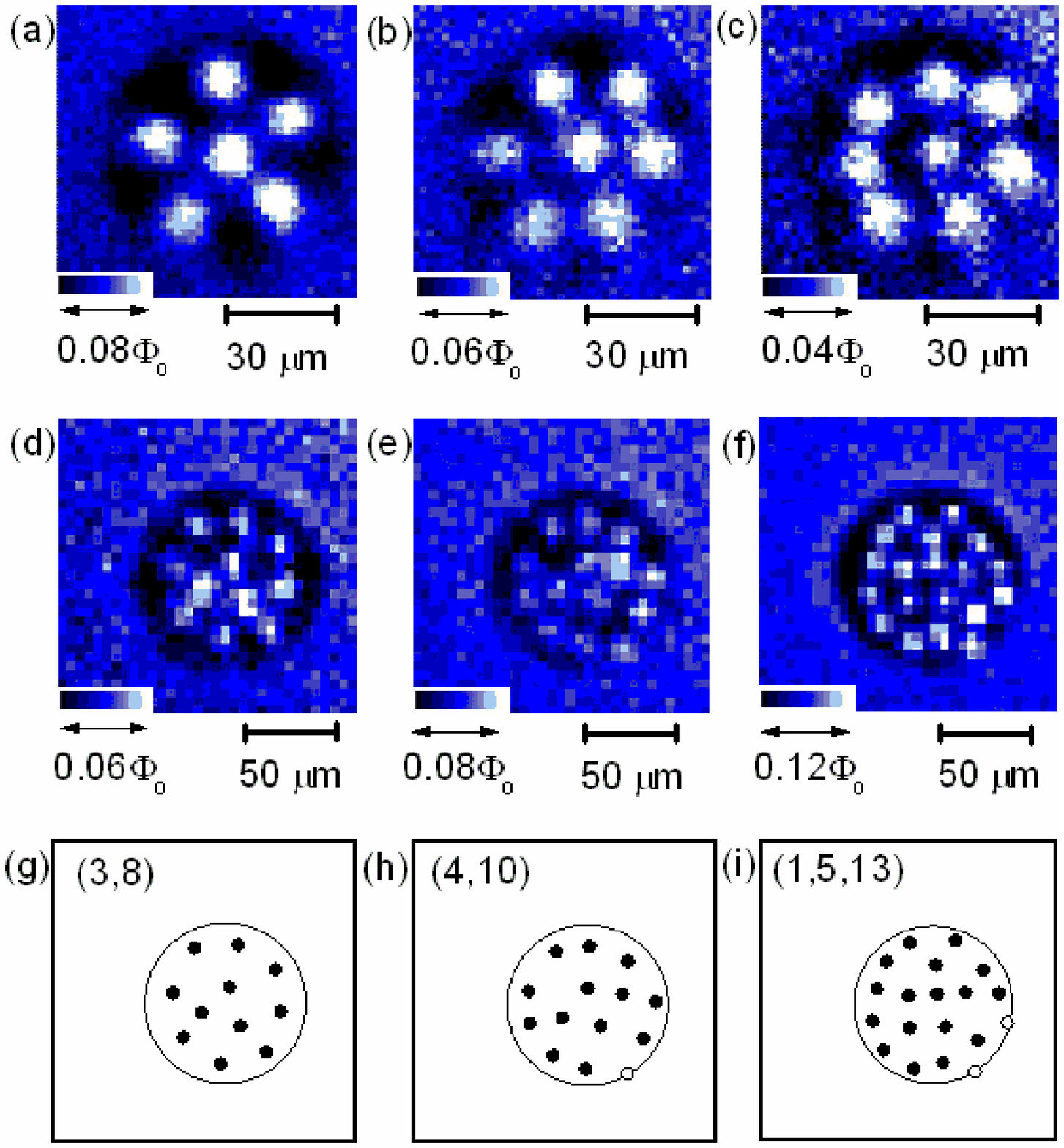, width=8cm} \vspace{0.1cm} \caption{
Images of vortices in disks with different diameters for large
vorticities. Images of (a), (b) and (c) are obtained in a 56 $\mu$m
disk in 3.9 K at fields of 8.0 $\mu$T, 9.0 $\mu$T, and 10 $\mu$T,
respectively. Other images of (d), (e), and (f) are observed in a
106 $\mu$m disk in 5.0 K at fields of 3.8 $\mu$T, 4.6 $\mu$T, and
6.0 $\mu$T, respectively. For clarity, the corresponding vortex
configurations for these images of (d), (e), and (f) are sketched in
(g), (h), and (i), respectively. }
\end{center}
\end{figure}

The growth of the inner shell occurs at even larger $L$. Fig. 4
shows the field evolution of vorticity, together with vortex
configuration, observed in the largest disk with 106 $\mu$m
diameter. The vortex shells with a single vortex at the center are
observed for $L$= 6, 7, and 8. This is consistent with the
observation made on the 56 $\mu$m disk discussed above. At $L$= 9 a
vortex pair appears as the inner shell, while the others form the
outer shell ring around, resulting in a (2,7) configuration. The
pair configuration of the inner shell is also observed at $L$= 10.
At $L$= 11 three vortices form a triangle as the inner shell,
resulting in a (3,8) configuration. The triangular inner shell is
observed over some field range and persists up to $L$= 13. In
contrast, the appearance of 4 vortices as the inner shell occurs
only at $L$= 14. At $L$= 15 five vortices form a pentagon as the
inner shell. The vortex shells with the pentagon inner ring appear
in vorticities up to $L$= 17 above which three concentric shells are
formed.

\begin{figure}
\begin{center}
\epsfig{file=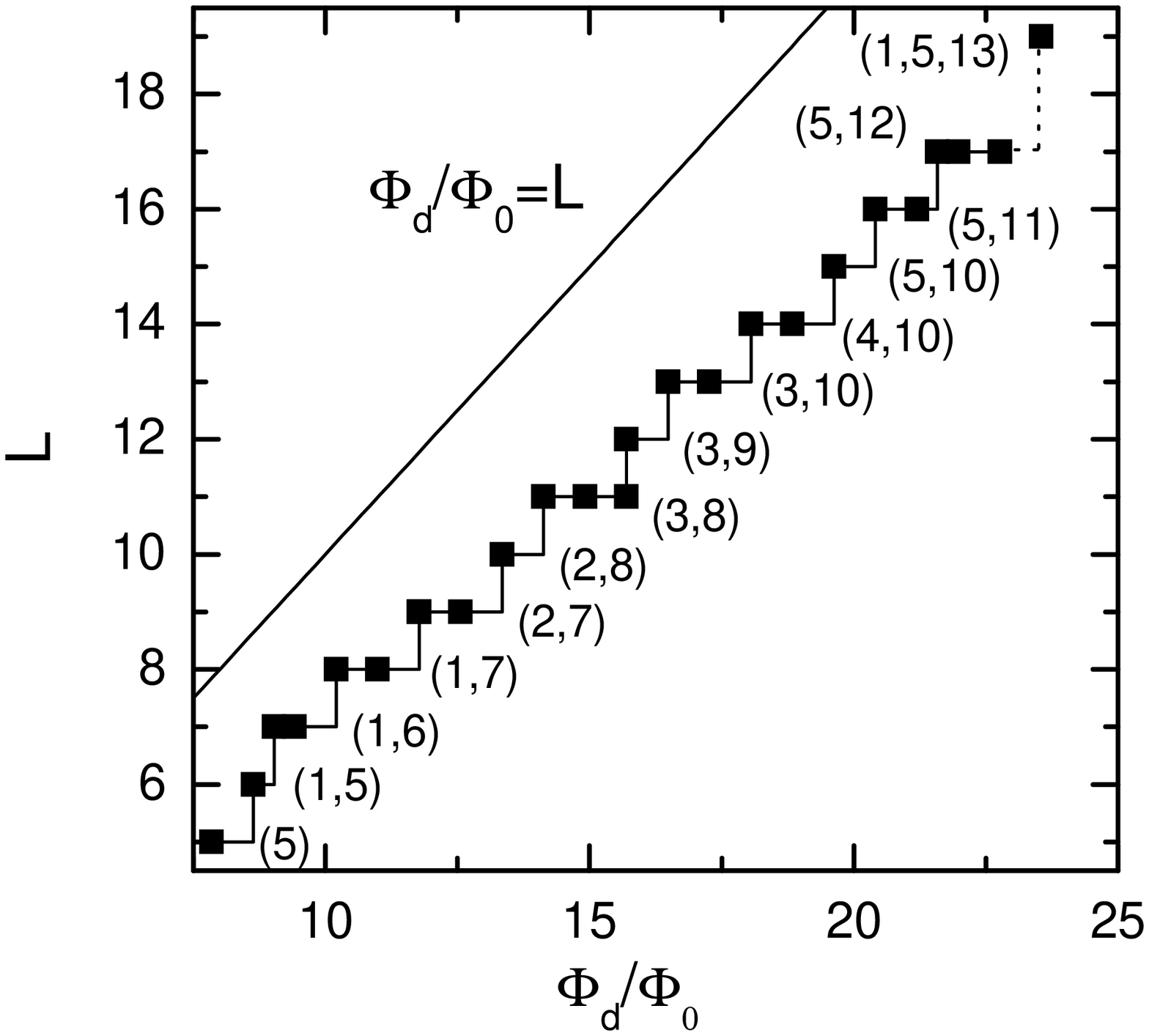, width=8cm} \vspace{0.1cm} \caption{ Vorticity
vs. normalized magnetic flux in the 106 $\mu$m disk in 5.0 K. The
configurations of vortices are denoted. A solid line represents a
condition of $\Phi_d/\Phi_0=L$.}
\end{center}
\end{figure}

These results follow mostly the rule of the shell filling obtained
in theoretical and numerical studies
\cite{BaelusPRB2004}\cite{MiskoPRB2007}. This includes the magic
number configuration at $L$= 6 where the first formation of two
concentric shells occurs, and also the formation of (2,7) at $L$= 9
where two configurations of (1,8) and (2,7) are energetically
comparable but the latter one becomes the ground state for the large
disk \cite{MiskoPRB2007}. However, some differences are found for
large vorticities ($\geq 13 $). The two shell configurations with 4
vortices in the inner ring are shown to appear for 13 $\leq L \leq$
15, while we observe only the (4,10) configuration at $L=14$.
Moreover, after the pentagon inner shell at $L=$ 16 the formation of
three vortex shells is shown to occur at $L=$ 17, which is not
observed above.

A physical origin for the differences would be found in actual
vortex configurations observed in the largest disk. Selected images
are shown in Figs. 3(d)-3(f). For clarity, the corresponding vortex
configurations are sketched in Figs. 3(g)-3(i). Although some
distortions are present, one can reasonably find quasi-symmetric
shell formation of vortices even in the largest disk. This is
exemplified by Figs. 3(d) and 3(g), where three vortices form a
triangle in the inner shell surrounded by 8 vortices as the outer
shell ring around. Focusing on vortex configurations in the inner
shell, we find the formation of a pair, the triangle and a pentagon,
of which patterns are similar to those observed for small
vorticities ($L \leq$ 5) given in Fig. 1. However, this is not the
case for a square. As shown in Figs. 3(e) and 3(h), the vortices in
the inner shell form a strongly deformed square like a rhombus. As
discussed below, this originates from the effect of the bulk pinning
quenched on the $\alpha$-MoGe film.


We often observe a vortex (or vortices) close to the rim of the disk
for large vorticities, which could be attributed to possible
roughness of the disk boundary. This is visible in the
configurations of (4,10) (Figs. 3(e) and 3(h)) and (1,5,13) (Figs.
3(f) and 3(i)).  As denoted by an open circle(s) in the
corresponding sketch, a vortex (vortices) appears close to the
lower-right rim of the disk. More rim vortices are occasionally
observed when the field-cool procedure is repeated in the same
magnetic field, indicating extra vortices trapped in the rim during
the field-cool process. If we neglect the rim vortex (vortices), the
differences between the experimental observation and the theoretical
prediction in vortex configurations found for $L \geq 13 $ would be
partly explained. In Fig. 3(h), the rim vortex becomes a part of the
outer shell ring. If we neglect the vortex, the configuration
becomes (4,9). This is consistent with the first configuration for 4
vortices in the inner shell given theoretically
\cite{BaelusPRB2004}\cite{MiskoPRB2007}. Similarity is applicable to
the (1,5,13) configuration shown in Fig. 3(i), where two rim
vortices are recognized. Thus, the configuration could be reduced to
(1,5,11), which is consistent with the magic number configuration
for three vortex shells \cite{BaelusPRB2004}\cite{MiskoPRB2007}.
Further studies would be needed to clarify the origin and the role
of the rim vortices on the vortex configurations in details.

Repeating field-cool measurements over some field range, we observe
configurations with different vorticities at the same magnetic
fields \cite{repeatability}. An example obtained on the 34 $\mu$m
disk is given to the inset to Fig. 2(a). Although we take the same
procedure of the field cooling, both the $L$= 5 polygon and $L$=
(1,5) shell are observed at the fields of $\Phi_{d}/\Phi_0=$ 10.4
and 10.9, resulting in some field crossover in the transition
between the $L$= 5 and $L$= (1,5) states. Similar crossover may be
present in every step in the stair-case behavior of $L$ given to
Fig. 2(a) and Fig. 4.


\section*{V. Influence of a pinning site upon vortex configurations}

Out of over 20 disks measured in this study, we found that some
disks have a relatively strong pinning site(s) \cite{pinning site},
although no defect is visible under an optical microscope. The
pinning center(s) is probably introduced during the sample
preparation and it originates from the slight compositional
inhomogeneity or subtle roughness in the disk. The influence of the
pinning site(s) on vortex configurations depends highly on where it
is present in the disk. As an example, we present the results (for
$L$ up to 13) obtained on a 56 $\mu$m disk with a pinning site near
the disk center.

Figs. 5(a)-5(i) show selected vortex images observed in the disk. We
find the pinning site from the image at $L$= 1 shown in Fig. 5(a),
where a vortex is unnaturally off-centered in the disk \cite{pinning
site2}. This is against the geometrical confinement and it should be
attributed to the influence of the pinning center. Thus, we identify
the vortex position as the pinning site and mark a red dot on the
image. No more pinning center is found in the disk.

The rest of the images indicate clearly how other vortices are
distributed in the presence of the vortex trapped at the pinning
site as marked by the red dot. Fig. 5(b) shows the image of two
vortices in the disk: One of them sits at the pinning site, while
the other appears in the opposite side with respect to the disk
center. We do not observe any orientational changes in the pair even
when the magnetic field is varied. Similarity is applicable to the
image of a triangle configuration given to Fig. 5 (c). Thus, the
influence of the pinning site fixes the vortex patterns and it
breaks the rotational degree of freedom in the vortex
configurations.

The deformation of vortex polygons occurs for vorticities more than
4. Fig. 5(d) shows the image of 4 vortices in the disk. The vortices
form a rhombus-shaped configuration, where the diagonal distance
between the pinned vortex and the diagonal vortex is shorter than
that for the other two vortices. This is similar in shape to the
disordered square in the inner shell observed in the largest disk
(see Figs. 3(e) and 3(h)).

A unique configuration of vortices is observed in the image at $L$=
6. As shown in Fig. 5(f), one can see two rows of three vortices
like a dice pattern of six with the pinned vortex in the middle of
the upper row. This pattern is stable against some field change and
no shell pattern is observed at $L$= 6. A precursor of the formation
of the rows can be seen in the image at $L$= 5 (Fig. 5(e)), where
the vortices form a pentagon with deformation inward by the pinned
vortex (the top of the pentagon) and  higher right three vortices
try to be aligned. The formation of the (upper) row remains in the
image at $L$= 7 (Fig. 5(g)). Thus, the pinning site near the disk
center triggers off the formation of the vortex row and stabilizes
the dice like vortex pattern at $L$= 6.

Because of the dice pattern of vortices at $L$= 6, the first
formation of two concentric shells occurs at $L$= 7.  This is larger
than $L$= 6 where the magic number configuration for two concentric
shells is observed in the "pin-free" disks. Thus, the shell filling
rule is altered by the pinning site near the disk center.

For $L>$ 7, the vortex shells are observed with the pinned vortex
being a part of the inner shell ring. As in the image of Fig. 5(h),
the pinned vortex is involved in the formation of a pair as the
inner shell, while the other vortices form the outer vortex ring
around. The corresponding sketch given in Fig. 5(k) allows us to
identify the configuration as (2,6).  To our knowledge, this
configuration has not been reported in other studies so far. For
larger vorticities, we observe two shell configurations of (2,7),
(2,8) and (3,8) which are similar to those observed in the largest
disk.

\begin{figure}
\begin{center}
\epsfig{file=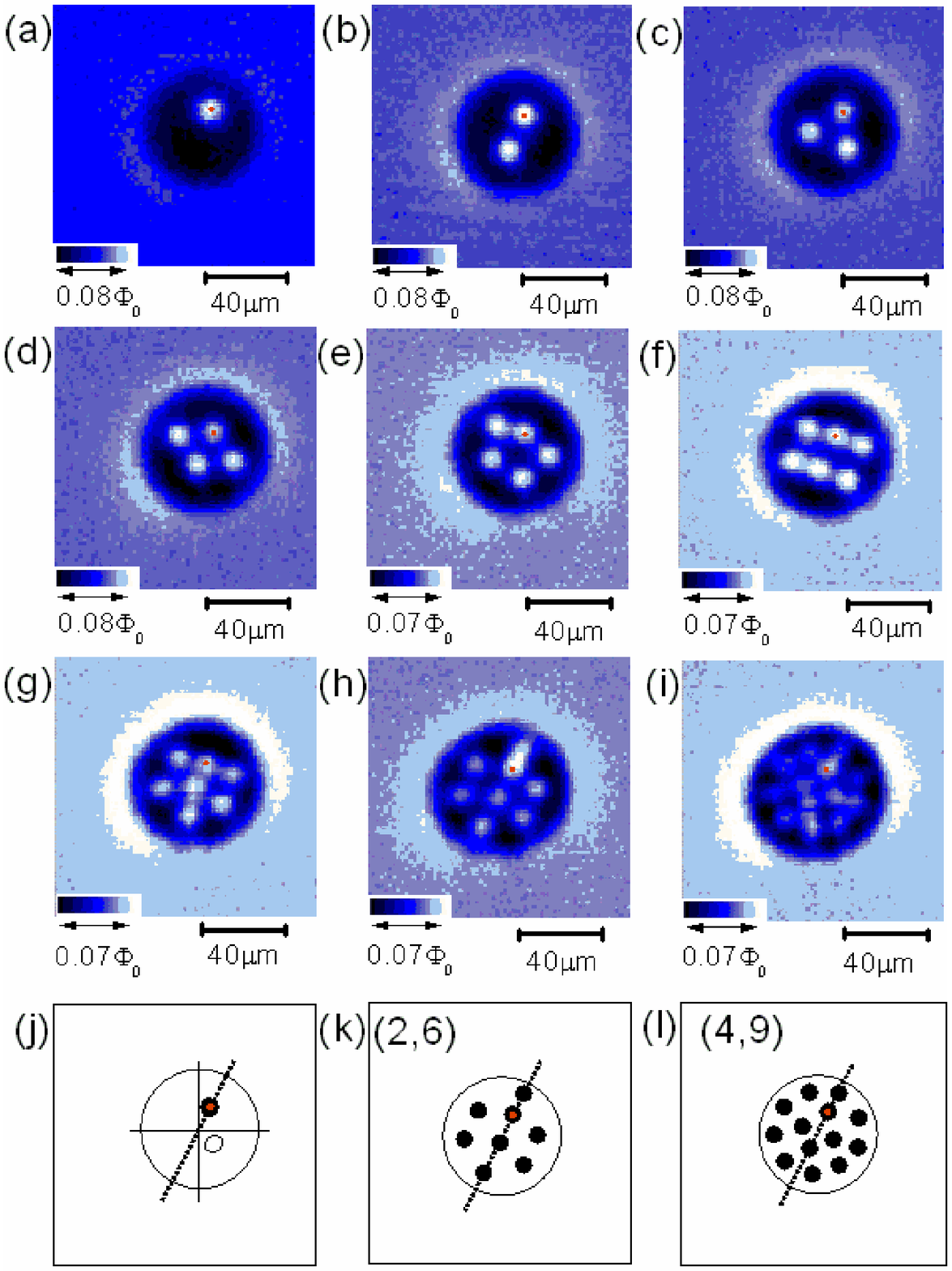, width=8cm} \vspace{0.1cm} \caption{
Selected images of vortices observed in a 56 $\mu$m disk with a
pinning center taken in 3.3 K for different vorticities. (a) $L$= 1
at 4.0 $\mu$T; (b)  $L$= 2 at 6.0 $\mu$T; (c) $L$= 3 at 7.5 $\mu$T;
(d) $L$= 4 at 9.0 $\mu$T; (e) $L$= 5 at 10 $\mu$T; (f) $L$= 6 at 11
$\mu$T; (g) $L$= 7 at 13 $\mu$T; (h) $L$= 8 at 13.5 $\mu$T; (i) $L$=
13 at 20 $\mu$T. A red point in each image marks the position of the
pinning site in the disk. For clarity, the vortex configurations for
the images of (a), (h), and (i) are sketched in (j), (k), and (l),
respectively. The dotted line for each sketch represents a symmetric
line defined by the pinning site and the disk center \textbf{O}.}
\end{center}
\end{figure}

It is interesting to show the image at $L$= 13 given in Fig. 5(i),
where 4 vortices form the inner shell surrounded by 9 vortices as
the outer shell. As sketched in Fig. 5(l), the vortices in the inner
shell form a rhombus-shaped, deformed square, while the outer shell
vortices form a ring around. Again, focusing on the inner shell, we
find that the deformation from the regular square is weaker than
that observed at $L$= 4 (Fig. 5(d)) and the effect of the pinning
site is opposite; Namely, the pinned vortex and the diagonal vortex
gives the longer diagonal of the rhombus. Thus, the vortex square is
stretched by the influence of the pinning site. We note that this
effect depends highly on where the pinning site is present and how
large the square polygon is. The regular square may appear if the
pinning site is located on the circumference of the ring defined by
the regular square vortices.

All the vortex configurations observed in the disk can be
characterized by a line symmetry, defined by the pinned vortex and
the disk center \textbf{O}, as sketched in Fig. 5(j). Drawing the
line in the sketch of Fig. 5(l), for instance, one can find that the
vortex configuration is divided into halves that are nearly the
mirror configurations of each other. Similarity is applicable to the
other configurations including unique ones at $L$= 6 and 8, although
slight deviation is present at $L$= 5 and 7. Note that the odd
(even) number of vortices lies on the line when $L$ is odd (even).
In the "pin free" disks no unique line symmetry characterizing all
the vortex patterns with different vorticities appears. Thus, we are
convinced that the line symmetry in the vortex configurations is
induced by the off-centered pinning site in the disk. A physical
reason for the symmetry is not simple since the vortex patterns are
determined by subtle balance of many competing interactions,
including the interaction with the bulk pinning, the repulsive
interaction between vortices, and the interaction with the disk
boundary. To clarify this issue, a numerical simulation study would
be needed.

Let us estimate roughly the pinning force $f_p$ for the vortex at
the pinning site. We find that at 4.5 $\mu$T, just below the step
field between $L=$ 1 and $L=$ 2 states, the disk has only a single
vortex, but it appears at the disk center.  This is distinct from
the observation made at lower fields of 3.5 $\mu$T and 4.0 $\mu$T,
where the vortex appears at the off-centered pinning site (Fig.
5(a)). These facts indicate the important influence of the force on
the vortex due to the geometrical confinement, and suggest that this
force should be comparable to the pinning force in between 4.0 $\mu$
T and 4.5 $\mu$ T. A model described in Refs. \cite{BuzdinPL1994},
\cite{BaelusPRB2004} and \cite{GrigorievaPRL2007} shows that the
force $f^s$ of vortex interactions with the shielding current and
edge for a given magnetic field is written as
\begin{equation}
 f^s= f_0 \Big[\frac{1}{1-(\frac{r}{D_d/2})^2}-\frac{\Phi_{d}}{\Phi_0}\Big]\cdot\frac{r}{D_d/2}
\end{equation}
where $r$ is the distance of the vortex position from the disk
center and
$f_0(=\Phi_0^2/\pi\mu_0D_d\lambda^2\approx5.0\times10^{-8}$
 N/m) is the unit of the force for a vortex per unit length. Using
the distance ($r=$ 12 $\mu$m) for the off-centered vortex, we find
the pinning force as $f_p \approx f^s \approx$ 1.8 $f_0$. This
result, together with our observation where the pinning site
accommodates only one vortex, are consistent with the numerical
simulation made at the weak pinning force of $f_p/f_0=$ 2
\cite{GrigorievaPRL2007}.


Further measurements are made on other disks with a pinning site
near the disk edge. It turns out that the pinned vortex near the
disk edge gives only some distortion on the vortex ring(s). No
unique configuration is observed except for a highly off-centered
vortex configuration at $L$= 1. These facts indicate that the
pinning site near the disk center provides more strong influence
upon the vortex configurations than that near the disk edge.

\section*{VI. Comparison with other experiment}

Comparing the results in the present experiment to those found in
the Bitter decoration experiment \cite{GrigorievaPRL2006}, we find
that both experiments share the general features of vortex
configurations including the shell filling rule and the magic
number, although magnetic fields and disk sizes are different in
decades. Let us shortly discuss the experimental conditions made on
these studies. In the decoration experiment the symmetric vortex
shells were observed in smaller disks with 1-5 $\mu$m diameter at
higher magnetic fields ($\sim$ mT). Due to the small disk size, the
geometrical confinement is strong and it dominates over the
influence of the strong bulk pinning in the Nb film. Meanwhile, in
the present experiment, the disk size is $\sim$ 10 times larger and
the field range is two decades smaller. The corresponding average
vortex spacing is $\sim$ 10 $\mu$m, which is larger than the
effective penetration depth. Thus, the interaction between vortices
as well as the influence of the edge current are weaker, but they
remain effective in the vortex configurations because the bulk
pinning in the amorphous film is weak. This situation is validated
in the observation of a reasonably ordered configuration of vortices
in $\alpha$-MoGe thin films at a low field of 10
$\mu$T\cite{NishioKokuboPRB2008}. Moreover, normalizing the disk
size by the effective penetration depth, we find that our disk size
(6-35 $\Lambda(0)$) is not much different from that of 9-46
$\Lambda(0)$ for the decoration experiment. These facts not only
link two experiments made in different sizes and different field
ranges, but also confirm that our observation is made in the large
\emph{mesoscopic} regime.

\section*{VII. Summary}
We presented the scanning SQUID microscope images of vortices
observed in the weak pinning, $\alpha$-MoGe thin disks with
different diameters. The observed images illustrate clearly
geometry-induced, (quasi-)symmetric configurations of vortices for
$L$ up to 19.

For small vorticities up to 5, the vortices form the symmetric
polygons with respect to the disk center. The polygon size grows
with $L$, while it shrinks by the enhancement of the screening
current. Some of the polygons rotate clearly with respect to the
disk center, indicating the rotational degree of freedom in the
vortex configurations. The influence of the geometrical confinement
is evident in the quantitative agreements between the field
evolution of the polygon size and the theoretical considerations in
the London limit.

For large vorticities more than 6, the vortices form the concentric
shell rings. The field evolutions of the inner and outer shell rings
follow the rule of the shell filling obtained in the theoretical and
numerical studies. Some differences are found in vortex
configurations for large vorticities ($\geq$ 13) observed in the
largest disk, and they could be explained partly by the influence of
the weak bulk pinning and/or the edge roughness in the disk.

We also presented the vortex images obtained in the disk with the
off-centered pinning site. The results reveal the pinning induced,
novel configurations of vortices: In addition to the deformed
polygons, we observed unique vortex configurations of the dice
pattern at $L$= 6 and (2,6) at $L$= 8. These lead to the pinning
induced changes in the shell filling rule and the magic number. All
the configurations observed in the disk are nearly symmetric with
respect to the line defined by the pinned vortex and the disk
center.

\section*{Acknowledgements}
N. K. thanks O. Narikiyo, H. Kawai, K. Kadowaki, B. J. Baelus, and
V. R. Misko for useful discussion. This work was supported partly by
the grant in Aid for Scientific research from MEXT (the Ministry of
Education, Culture, Sports, Science and Technology), Japan.

{Corresponding author: N. Kokubo, Center for Research and
Advancement in Higher Education, Kyushu University, 4-2-1, Moto-oka,
Nishi-ku, Fukuoka, Fukuoka 819-0395, Japan;
e-mail:kokubo(at)rche.kyushu-u.ac.jp.}

\end{multicols}{2}


\begin{thebibliography}{00}
\bibitem{rmsnoise} The rms flux noise for our system referred to the pick-up loop is $\sim$ 5
$m\Phi_0$ per pixel in the images.
\bibitem{rotation} We note that the observed orientations are some of
the metastable states stabilized by subtle uneven boundary
confinement due to the edge roughness and/or possible weak bulk
pinning in the disk. Otherwise, during the scanning, the orientation
can be aligned along the scan direction of the pick-up loop due to
the weak coupling between the loop and the vortices.
\bibitem{repeatability} Repeated
field-cool measurements at the same magnetic fields show nearly
identical vortex configurations, including the orientation of the
patterns, when the vorticity is unchanged.
\bibitem{pinning site2} The pinning site at the disk center can be
found from a vortex pattern at $L$=2.
\end{thebibliography}
\end{document}